\documentclass[twocolumn,letter]{jpsj2}
\title
{Quantum-Classical Transition in Dissipative Double-Well Systems\\
---A Numerical Study by a New Monte Carlo Scheme---}
\author{\textsc{Takeshi MATSUO}$^{1,2}$%
\thanks{E-mail address: matsuo@phys8.s.chiba-u.ac.jp}, %
\textsc{Yuhei NATSUME}$^1$ and \textsc{Takeo KATO}$^2$%
\thanks{E-mail address: kato@issp.u-tokyo.ac.jp}, %
}
\inst{${\ }^1$ Graduate School of Science and Technology, Chiba University,
Inage-ku, Chiba 263-8522 \\
${\ }^2$Institute for Solid State Physics, University of Tokyo,
Kashiwa, Chiba 277-8581}

\abst{The thermodynamics of dissipative quantum systems with double-well potentials
is studied by a path-integral Monte Carlo (PIMC) method without truncation to
the two-state model. For efficient simulation at low temperatures,
we develop a new local update scheme on the basis of the approximate decomposition
of the Boltzmann weight to Gaussian distributions.
The localization transition induced by ohmic dissipation is clarified numerically
for arbitrary potential barriers.
}
\kword{Caldeira-Leggett model, quantum dissipation, path-integral Monte Carlo,
localization transition}
\begin{document}
%~
%\newpage
\maketitle
Quantum tunneling between distinct states subject to dissipation
has been one of the central problems
in both physics and chemistry for several decades.~\cite{Leggett87, Weiss99}
A number of studies on dissipative quantum systems have revealed that
dissipation strongly affects the quantum nature of the system.
In particular, the suppression/destruction of quantum superpositions
has been studied extensively in the dissipative two-state model.
The most striking feature of the two-state system is that
localization transition occurs at a critical value of
dissipation strength for ohmic damping.~\cite{Chakravarty82, Bray82, Schmid83}
This phenomenon can be interpreted as a `quantum-classical transition' driven
by the coupling with the environment, which is unavoidable for macroscopic objects.

The dissipative two-state model is useful for the analysis of generic systems
with a double-well-shaped potential in the large-barrier limit.
For the small-barrier case, however,
the truncation to the two-state model cannot be justified, and we must
analyze the original generic system directly. Dissipation effects on generic
systems are nontrivial, and have been considered only in a few studies.
Aoki and Horikoshi have discussed
localization transition in the double-well potential systems by a nonperturbative
renormalization group approach with local approximation.~\cite{Aoki02}
They concluded that the critical value of
dissipation strength is larger than that of the two-state model.
Capriotti~{\it et al.} have studied the thermodynamics of dissipative double-well
systems by the path-integral Monte Carlo~(PIMC) method.~\cite{Capriotti02}
However, they have not discussed localization transition, because their algorithm
based on the conventional update scheme significantly limits the accuracy of
the simulation at low temperatures.
Thus, neither the identification of localization transition nor the determination
of critical dissipation strength have been achieved so far without approximations.

Recently, a significant improvement in the update scheme of the PIMC method
has been presented by Werner and Troyer
for periodic-potential systems.~\cite{Werner05a,Werner05b}
They introduced a novel update scheme by extending the cluster algorithm
for classical Ising-spin systems~\cite{Swendsen87, Wolff89, Luijten95}
to realize efficient change in the paths beyond the potential barrier.
In addition, they adopted an efficient local update scheme by combining the
fast random-number generation of the Gaussian distributions with
stochastic acceptance.
They demonstrated that their algorithm is much more efficient than
conventional PIMC simulations,~\cite{Herrero02,Kimura04} and
determined the phase diagram successfully by the accurate calculation of
response functions at low temperatures.
This development of the Monte Carlo algorithm, in principle, enables us to study
the thermodynamics of various types of dissipative systems in much wider
parameter regions.
The application of their method to double-well systems,
however, is not straightforward. The cluster update can be implemented successfully
to double-well systems, whereas their local update scheme
has difficulty in implementing low acceptance rates at low temperatures.
This is because their local update scheme is based on the fact that
a particle in a weak periodic potential can move almost freely, whereas
a particle in a double-well potential is trapped in the whole
well-shaped structure, even for a small potential barrier.

In this paper, we propose a new scheme for efficient local update
in the PIMC simulation, and demonstrate its efficiency for large system sizes.
By combining this improved update with the cluster update by Werner and Troyer,
we show for the first time the quantum-classical transition in generic
double-well potential systems without truncation to the two-state model.
We stress that our local update scheme has wide application to various classes
of models and is not restricted to dissipative systems.

The dissipative system is described by the Caldeira-Leggett
model:~\cite{Feynman69, Leggett81}
\begin{eqnarray}
& & H = \frac{p^2}{2m} + V(x) \nonumber \\
& & + \sum_\alpha \left\{ \frac{p_\alpha^2}{2m_\alpha} +
\frac{m_\alpha\omega_\alpha^2}{2} \left( y_\alpha -
\frac{c_\alpha} {m_\alpha\omega_\alpha^2} x \right)^2 \right\}.
\end{eqnarray}
The environment properties are uniquely determined by the spectral function
\begin{equation}
J(\omega) = \frac{\pi}{2} \sum_\alpha \frac{c_\alpha^2}{m_\alpha\omega_\alpha}
\delta(\omega-\omega_\alpha).
\end{equation}
In this paper, we restrict our discussion to ohmic dissipation, for which
the spectral function is given as $J(\omega) = m \gamma \omega$.
The double-well potential is expressed by the quadratic-plus-quartic form
\begin{equation}
V(x) = \frac{16 V_0}{a^4} \left( x^2 - \frac{a^2}{4} \right)^2,
\end{equation}
for a barrier height $V_0$ and a spatial distance $a$ between
two potential minima.
In the following discussion, we rescale the length
by $a/2$, the energy by $V_0$ and the frequency
by $\omega_0 = (32V_0/ma^2)^{1/2}$ for
a small oscillation near the potential minima.
Then, in the absence of dissipation, the system
can be characterized by only one parameter $g = \hbar \omega_0/V_0$.
The limit $g \to 0$ corresponds to the semi-classical limit,
in which the system can be truncated to the two-state model using
the dilute instanton approximation.~\cite{Leggett87} We further define a
dimensionless dissipation strength as
$\alpha = m\gamma a^2/(2\pi \hbar)$.
By integrating the variables of the harmonic oscillators, we can rewrite
the partition function in terms of path integrals using the effective action as
\begin{eqnarray}
& & \hspace{-3mm} Z = \int {\cal D} x(\tau) \exp\left( - S[x(\tau)] \right), \\
& & \hspace{-3mm} S[x(\tau)] =  \frac{8}{g} \int_0^{\beta g} \! {\rm d}\tau
\left\{ \frac{1}{2} \left(\frac{{\rm d}x}{{\rm d}\tau}\right)^2
+ \frac{1}{8} (x^2 - 1)^2 \right\} \nonumber \\
& &  - \int_0^{\beta g} \! {\rm d}\tau \int_0^{\beta g} \! 
{\rm d}\tau' \, k(\tau - \tau') x(\tau) x(\tau'),
\end{eqnarray}
where the memory kernel $k(\tau)$ is given as
\begin{equation}
k(\tau)
=
\frac{\alpha}{4} \frac{(\pi/\beta g)^2}{\sin^2\left( (\pi/\beta g) \tau \right)}.
\end{equation}

For the PIMC simulation, the path is discretized by the Trotter number $N$,
which is taken to be sufficiently large to
satisfy $\Delta \tau \equiv \beta/N \lesssim 1$.
After the discretization, the action is obtained by
the path fragment $x_j = x((\beta g /N)j)$ as
\begin{equation}
S[\{x_j\}] = - \sum_{i, j} K(i-j) x_i x_j + \Delta \tau \sum_j (x_j^2 -  1)^2,
\label{DiscreteAction}
\end{equation}
where the long-range kernel including the kinetic energy is given by
\begin{equation}
K(i) = \frac{4N}{\beta g^2} (\delta_{i,1} + \delta_{i,N-1}) 
+ \frac{\alpha}{4} \frac{(\pi/N)^2}{\sin^2((\pi/N) i)}, 
\end{equation}
for $i\ne 0$. The value of $K(0)$ is taken so that 
the zero-frequency Fourier component of the kernel becomes zero.
In the conventional update scheme, one of the path fragments is changed as
$x_j^{{\rm new}} = x_j^{{\rm old}} + \varepsilon$ for a fixed $j$.
Then, this change is accepted
with the probability $\min(\exp( - (S[\{x_j^{{\rm new}}\}] - S[\{x_j^{{\rm old}}\}] )),1)$.
Here, the displacement $\varepsilon$ is generated randomly by
the uniform distribution with the range $[ - \delta, \delta]$,
where $\delta$ is chosen appropriately to obtain the optimal acceptance rate.
This algorithm, however, takes considerable time for one sweep of local updates
on all the path fragments.

\begin{figure}[tb]
\begin{center}
\includegraphics[width=80mm]{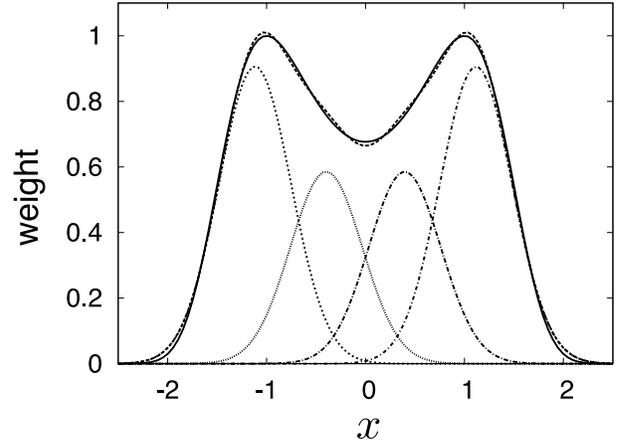}
\caption{Boltzmann weight as a function of position variable $x$
for double-well potential (solid line) and
approximated potential (broken line), which is a sum of
four Gaussian distributions (other lines).}
\label{fig:potential}
\end{center}
\vspace{-4.5mm}
\end{figure}
We now introduce a new update scheme. First, the Boltzmann weight of the
potential term is approximately decomposed into $M$ pieces of
the Gaussian distribution as
\begin{equation}
{\rm e}^{-\Delta \tau V(x)}
\simeq
\sum_{m=1}^{M} \mu_{m}\, {\rm e}^{-\lambda(x-\sigma_{m})^2},
\label{decomposition}
\end{equation}
where $\lambda$, $\sigma_m$ and $\mu_m$ are determined by fitting.
For the double-well potentials, $V(x) = (1-x^2)^2$, and
this approximate decomposition can be carried out with a small value of $M$.
One example of fitting is shown
in Fig.~\ref{fig:potential} for $\Delta \tau = 100/256 \simeq 0.4$.
Here, we used $M=4$ Gaussian distributions, and the parameters were obtained as
$\lambda = 3.7683$, $- \sigma_1 = \sigma_4 = 0.9068$,
$-\sigma_2 = \sigma_3 = 0.5858$,
$\mu_1 = \mu_4 = 1.1169$ and $\mu_2 = \mu_3 = 0.3963$.
As seen in Fig.~\ref{fig:potential},
the approximate fitting is almost in agreement with the original weight.
In our calculation, we checked that the deviation from the exact weight is
not distinguishably different for the obtained results up to $N=1024$.
By this decomposition, the partition function is written in the form
\begin{equation}
Z = \sum_{\{ m_j \}} \int \prod_j {\rm d}x_j \, W(\{x_j\}, \{m_j\}).
\end{equation}
Here, the new variables $\{ m_j \}$ play the role of auxiliary fields.
Our update scheme consists of two steps.
In the first step, the variables $\{ m_j \}$ are chosen for fixed
$\{ x_j \}$ with the probability $t(m_j  | x_j )$.
For this update, the weight is modified as
$W(\{ x_j \}, \{ m_j \}) = f(\{ x_j \}) \times \prod_j w(x_j ,m_j)$, where
$w(x_j, m_j) = \mu_{m_j} \exp(-\lambda (x_j - \sigma_{m_j})^2)$
and $f(\{x_j\})$ is independent of $\{ m_j \}$.
Then, the transition probability is given by
the extended detailed valance condition~\cite{Kawashima04} as
\begin{equation}
t( m_j | x_j ) = \frac{w(x_j, m_j)}{\sum_{m_j}w(x_j,m_j)}.
\end{equation}
Note that this update needs only the local information of each path fragment,
thus, can be performed efficiently.
In the second step, the path fragments $\{ x_j \}$ are updated for
fixed $\{ m_j \}$.
In this update, it is crucial that for fixed $\{ m_j \}$,
the effective action defined by $W(\{ x_j \}, \{ m_j \}) = \exp( - \bar{S} )$
is written in a quadratic form of $\{ x_j \}$.
Using a Fourier transformation, the effective action can be written in
the normal form
\begin{eqnarray}
& & \hspace{-10mm}
\bar{S}
=
\sum_n \left( c_n | \tilde{x}_n |^2 + \lambda N |\tilde{x}_n - \tilde{\sigma}_n | ^2
\right) - \sum_j \ln \mu_{m_j}, \\
& & \hspace{-10mm}
c_n
=
\frac{16N^2}{\beta g^2} \sin^2 \frac{\pi n}{N} + \frac{\pi^2 \alpha}{2} |n|,
\end{eqnarray}
where $\tilde{x}_n$ and $\tilde{\sigma}_n$ are the Fourier components of $x_j$ and $\sigma_{m_j}$,
respectively. 
The effective action can be modified further as
\begin{equation}
\bar{S}= \sum_n (c_n + \lambda N) \left| \tilde{x}_n - \frac{\lambda N \tilde{\sigma}_n}{c_n + \lambda N}
\right|^2 + g(\{ m_j \}),
\label{barS}
\end{equation}
where $g(\{ m_j \})$ is independent of $\{ x_j \}$. Then, the new path is efficiently generated
by the Gaussian distribution with mean $\lambda N \tilde{\sigma}_n/(c_n + \lambda N)$
and standard deviation $(c_n + \lambda N)^{-1/2}/2$ for both real
and imaginary parts of $\tilde{x}_n$.~\cite{footnote1}
Note that all the path fragments are updated immediately without rejection.

In the presence of large potential barriers, global change of the path across
the barrier is strongly suppressed, even in the improved local update.
To overcome this difficulty, we adopt the cluster update by
Werner and Troyer.~\cite{Werner05a} We first focus on the
mirror symmetry $V(x) = V(-x)$ for the potential energy, and flip the sign of
the path without varying its amplitude, $|x_j|$.
Since the potential term in eq.~(\ref{DiscreteAction}) is unchanged
by this update of the path, we can disregard this term in the following discussion.
We define the Ising variables as $x_j = s_j |x_j|$,
and express the effective action by these variables as
\begin{equation}
S[\{s_j\}] = -\sum_{i,j} J_{i,j} s_i s_j,
\end{equation}
where $J_{i,j} = K(i-j) |x_i| \!\cdot\! |x_j|$.
This is simply the one-dimensional long-range Ising model.
Then, the Wolff cluster algorithms can be applied directly.~\cite{Luijten95}
We note that the cluster update can deal with quantum tunneling processes
efficiently through the creation (annihilation) of instantons
and the change in their positions.

\begin{figure}[tb]
\begin{center}
\includegraphics[width=80mm]{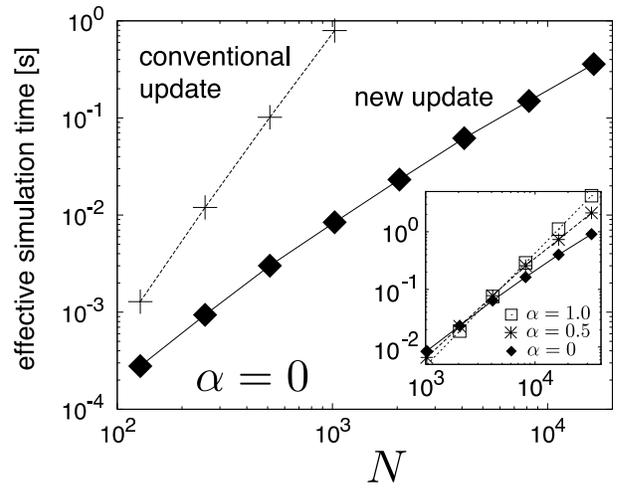}
\caption{Effective simulation time for Monte Carlo sampling during
autocorrelation time measured
in correlation function $\langle x(0) x(\beta/2)\rangle$.
Main panel: comparison between conventional and new local update scheme
with no cluster update for $\alpha = 0$ (no dissipation) and $g = 2$.
Inset: $\alpha$-dependence of effective simulation time
for both cluster update and new local update.}
\label{fig:potential:b}
\end{center}
\vspace{-4.0mm}
\end{figure}
We show the effectiveness of our improved local update in Fig.~\ref{fig:potential:b}.
Here, the CPU time in a improved calculation is measured in a cluster computer
consisting of eight Pentium 4 processors for the Monte Carlo sampling within
the autocorrelation time in
the correlation function $\langle x(0) x(\beta/2) \rangle$.
The main figure in Fig.~\ref{fig:potential:b} shows
the comparison between the conventional algorithm and the new algorithm for the
local update at $\alpha=0$ (no dissipation) without the cluster update.
As seen in this figure,
our algorithm for the local update shows significant improvement in
the simulation time.
The improved CPU time is approximately proportional to $N$ for large system sizes,
for which the autocorrelation time is almost independent of $N$.
This system-size dependence indicates that the efficiency of our simulation is
limited by the Fourier transformation,
which requires time on the order of $N \log N$.
The inset in Fig.~\ref{fig:potential:b}
shows the CPU time measured for $\alpha \ge 0$ (nonzero dissipation) using
both the new local update and the cluster update.
This result shows that the efficiency
of our algorithm is less affected by dissipation strength $\alpha$.

\begin{figure}[tb]
\begin{center}
\includegraphics[width=80mm]{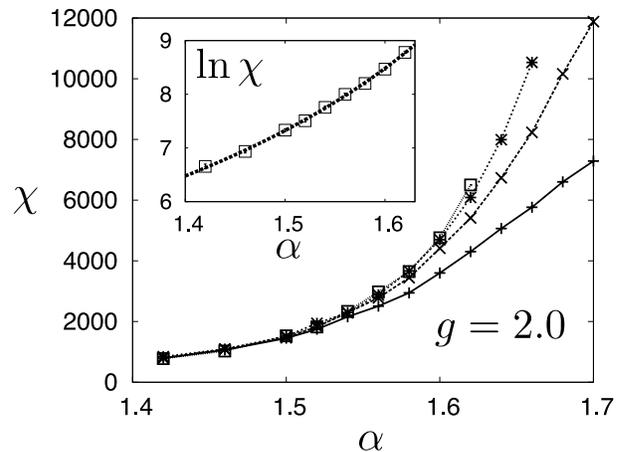}
\caption{Calculated susceptibility of double-well potential system as
a function of damping parameter $\alpha$. The curves correspond to
the Trotter numbers $N=2^{14},2^{15},2^{16}$ and $2^{17}$ from
the bottom to the top.
The inset shows the scaling fitting for $N=2^{17} \simeq 1.3\times 10^5$
in the region $1.42\le\alpha\le 1.62$.
}
\label{fig:chi}
\end{center}
\vspace{-4.8mm}
\end{figure}

Next, let us move onto the localization transition in double-well potential systems.
We expect that this transition belongs to the Kosterlitz-Thouless type because
of the same symmetry as the long-range Ising model with inverse square-law
interactions.~\cite{Luijten01}
In accordance with ref.~\citen{Luijten01}, the transition is
determined by calculating the uniform susceptibility defined as
$\chi  = N(\sum_{i,j} x_i x_j / N^2)  / ( \sum_i x_i^2 / N )$.
We use renormalization group analysis~\cite{Amit80} suggesting that
\begin{equation}
\chi
\sim
\exp [ A (\alpha - \alpha_{\rm c})^{-\nu} + B (\alpha - \alpha_{\rm c})^{\nu}
+ {\cal O}((\alpha-\alpha_{\rm c})^{\nu}) ].
\end{equation}
The critical value $\alpha_{\rm c}$ is then obtained by fitting.
We show the data of the susceptibility at $g=2.0$ for large system sizes up
to $N = 2^{17} \simeq 1.3 \times 10^5$ in the main panel of Fig.~\ref{fig:chi}.
To plot one point in this figure,
we used 64 independent runs, each of which consisted of 500 Monte Carlo steps (MCS)
for thermalization and next 600 MCS for measurement.
In all the runs, we checked that the autocorrelation time measured in the
correlation function $\langle x(0) x(\beta/2) \rangle$ is much less than 300 MCS.
The inset shows the fitting for $1.42 \le \alpha \le 1.62$,
where the data has almost converged on one curve with respect to
the change in the system size.
Stable fitting is obtained for each value of $g$, and the critical dissipation strength
$\alpha_{\rm c}$ is successfully determined within reasonable statistical errors.

\begin{figure}[tb]
\begin{center}
\includegraphics[width=80mm]{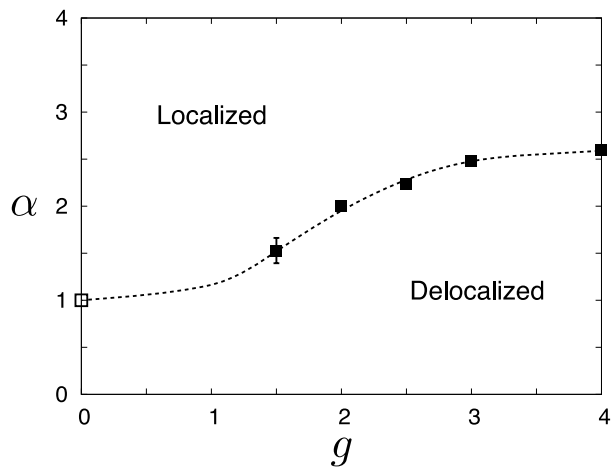}
\caption{Phase diagram of dissipative double-well systems
in $g$-$\alpha$ plane, where $\alpha$ is the dissipation strength and
$g=\hbar\omega_{0}/V_{0}$ is the zero-point quantum fluctuation normalized by
the potential barrier. The broken line is drawn to guide the eyes.}
\label{fig:phase}
\end{center}
\vspace{-4.8mm}
\end{figure}
Finally, the phase diagram is given in Fig.~\ref{fig:phase},
where the closed squares show the
critical values of dissipation strength for localization, $\alpha_{\rm c}$.
We stress that it is the first time that transition points in the present system
have been determined as a function of $g$ without approximation.~\cite{footnote2}
The error bar shown for the data at $g=1.5$ expresses the statistical error
estimated from fitting using five independent Monte Carlo results.
The error is reduced for larger values of $g$, and
becomes smaller than the symbol size at $g=4$.  As $g$ increases,
$\alpha_{\rm c}$ is enhanced from the value $\alpha_{\rm  c} = 1$, predicted
in the two-state limit $g \rightarrow 0$ (shown in the figure by the open square).
Furthermore, the tendency of saturation is clearly observed for larger values of $g$.
The behavior of $\alpha_{\rm c}$ indicates that
even for the infinitesimal potential barrier ($V_0 \ll \hbar \omega_0$),
a finite strength of dissipation is sufficient for the localization transition.
This impressive result has already obtained for
dissipative periodic-potential systems,
where the critical value of dissipation strength is obtained
as $\alpha_{\rm c}=1$ in the small-barrier limit
$V_0/\hbar \omega_0 \rightarrow 0$.~\cite{Schmid83,Fisher85} Note that
the periodic-potential system is special because of the existence of
a property called `duality', which strongly suggests that
the critical value $\alpha_{\rm c}$ is independent of the potential
barrier.~\cite{Fisher85}
Actually, for the large-barrier limit, the critical value is also given
as $\alpha_{\rm c}=1$. Because double-well potential systems do not have
such a property, the critical value can be derived from the limiting value.
%We stress that it is the first time to observe the deviation of $\alpha_{\rm c}$
%from the limiting value $\alpha_{\rm c} = 1$ in numerical simulations.

%The Gaussian decomposition technique (\ref{decomposition}) used in this paper
%has quite wide application.
The improved local update developed by us can be applied to other
models, one of which is the nonlinear harmonic oscillator system
such as the $\phi^4$ chain.
The present scheme is also useful for analyzing the nonlinear effects of
phonons in dielectric substances.
The extension of our algorithm to these systems will be studied elsewhere.
The idea of Gaussian decomposition also suggests that a model with continuous
variables can be mapped approximately into another model only with discrete
variables after integrating out the continuous variables.
This process will give a new insight into the original continuum model.

In summary, we considered the localization transition in
dissipative double-well systems by a PIMC simulation improved by
an approximate decomposition of the weight by several Gaussian distributions.
We determined the transition point by fitting
to the scaling expression for the Kosterlitz-Thouless transition.
For potential barriers that are small compared with quantum fluctuations,
the critical value of dissipation strength becomes larger than $1$,
and saturates at a finite value. We expect that the decomposition into
Gaussian distributions is helpful for both creating algorithms and
analyzing of other models with continuous variables.

We would like to thank N. Kimura for supporting preliminary calculations.
We also thank N. Kawashima for suggesting the application of the cluster
algorithm to the present system. The computation in
our work was carried out using the facilities of the Supercomputer Center,
Institute for Solid State Physics, University of Tokyo.
This work was partially supported by a Grant-in-Aid for Scientific Research from
the Ministry of Education, Culture, Sports, Science and Technology.


\begin{thebibliography}{99}
\bibitem{footnote1}Note that eq.~(\ref{barS}) is a somewhat abbreviated form.
For even $N$, the component $\tilde{x}_n$ ($n=0, N/2$) only has a real part.
So, for these components,
the standard deviation in the Gaussian random generator must be
taken as $(2(c_n + \lambda N))^{-1/2}$.
\bibitem{footnote2}In ref.~\citen{Aoki02}, the critical dissipation strength 
for large quantum fluctuations 
($g \ge 8$) is obtained by an approximate method as $\alpha_{\rm c} \sim 8$, which
is much larger than the value obtained by extrapolation of our results.
This discrepancy indicates the possibility that the local approximation adopted
in their analysis captures the localization transition only at a qualitative level.
\end{thebibliography}
\end{document}